\documentclass[letterpaper, 12pt]{article}%
\usepackage{graphicx}%
\usepackage[letterpaper, margin=1in]{geometry}
\usepackage{mathptmx}
\usepackage{mathtools}
\usepackage{amsmath}
\usepackage{amssymb}

\usepackage{subfig}
\usepackage{tikz} 
\usepackage{algorithm}
\usepackage[noend]{algpseudocode}

\usepackage{xcolor}

\usepackage{verbatim}
\usepackage{natbib}

\usepackage{mycommands}

\usepackage{setspace}
\usepackage[compact]{titlesec} 
\titlespacing{\section}{2pt}{2pt}{2pt} 
\usepackage{titling}
\title{Incorporating circuit theory into a dynamic model for crowd-sourced observations of migratory birds}
\author{Michael F. Christensen and Peter D. Hoff}
\date{Department of Statistical Science, Duke University \\
\today} 

\DeclareMathAlphabet{\mathbfit}{OML}{cmm}{b}{it}

\begin{document}
\setstretch{1.75}
\maketitle

\begin{abstract}
While the overarching pattern of biannual avian migration is well understood, there are significant questions pertaining to this phenomenon that invite further study. Necessary to any analysis of these questions is an understanding of how a given species' spatial distribution evolves in time. While studies of animal movement are often conducted using telemetry data, the collection of such data can be time- and resource-intensive, frequently resulting in small sample sizes. Ecological surveys of animal populations are also indicative of species distribution trends, but may be constrained to a limited spatial domain. Within this article we utilize crowd-sourced observations from the eBird database to model the abundance of migratory bird species in space and time. While crowd-sourced observations are individually less reliable than those produced by experts, the sheer size and spatial coverage of the eBird database make it attractive for use in this setting. We introduce a hidden Markov model for observed bird counts utilizing a novel transition structure developed using principles from circuit theory. After illustrating model properties we fit it to observations of Baltimore orioles and yellow-rumped warblers within the eastern United States and discuss insight it provides into the migratory patterns for these species.
\end{abstract}


\section{Introduction}

The migratory patterns for many bird species found within the northern hemisphere can be broadly characterized by northward flight to breeding grounds in the spring, and southward flight to wintering grounds during the fall. While the overarching patterns observed during bird migration are well documented, there is scientific interest in understanding how these migratory patterns are evolving, or may be expected to evolve over time \citep{barton2018}, in response to climate change \citep{visser2009} or loss of habitat \citep{sutherland1996}. There is also interest in developing methods to better identify species- and population-specific flyways \citep{buhnerkempe2016}, the routes used by migratory birds, and which tend to be characterized by environmental features conducive migration, such as abundance of food, water, and shelter and lack of mountain ranges or major shifts in elevation \citep{alerstam1993}. In addition to these questions, many of the mechanisms behind bird migration are poorly understood at present. There is for example little consensus regarding what actually triggers the onset of migration season for any given species, whether a response to shifting day lengths, local climate conditions or innate biological factors \citep{helbig2003, cornell2021}. 

Studies of avian migration generally require the collection of data on the location across time of individual birds or bird populations. Telemetry data, produced by capturing, tagging and monitoring the position of individual specimens in real time via satellite, GPS or radio \citep{dunn1977, perras2012}, is one of the most important data types for studies of animal movement \citep{hooten2017}. However, the collection of such data is time and resource intensive. and may often lead to small sample sizes and limited geographic domains \citep{hebblewhite2010}. It has also been noted that traditional telemetry data cannot be obtained for bird species which are too small to carry the tracking devices used to monitor location \citep{perras2012}. Professional monitoring surveys can also be a valuable tool for understanding species abundance within a studied region, \citep{lindenmayer2010} but once again, such data is both expensive and time consuming to obtain and provides only a snapshot of the species distribution during the time and within the vicinity that the survey was conducted \citep{caughlan2001}. Given that migratory birds may inhabit locations spanning thousands of kilometers during the course of a single year \citep{helbig2003}, individual monitoring surveys are fundamentally limited in their ability to inform scientists of larger-scale migratory patterns. Given the limitations of these data types, we wish to consider the feasibility and potential value in using crowdsourced observations in the study of bird migration.

\subsection{eBird data}
Bird watching as a hobby has seen considerable growth in recent decades \citep{schwoerer2022}. The eBird database is a project managed by the Cornell Lab of Ornithology that crowdsources observational data by allowing any user to record and submit their bird watching observations to the website. Observations may be reported simply by indicating that a bird from a given species was observed at the user's location, or more rigorously in the form of time- and location-indexed ``complete checklists" containing an exhaustive list of the counts and species of every bird observed by the submitting user, along with the length of time spent bird watching at that location \citep{ebird}. First launched in 2002 in collaboration with the Audubon society, the database has since recorded observations of over one billion birds and contains records from every country on earth \citep{ebird2024}. As the contributing user base and volume of freely available data has increased, so too has an awareness of the eBird database's value for scientific research. Numerous peer-reviewed studies \citep[e.g.][]{adde2021,tang2021,haas2022,bianchini2023,hochachka2023} of bird population and species distribution trends utilizing the eBird database have been conducted and published. While the amateur collection process and crowdsourced nature of these data has raised questions about their reliability and bias \citep{hochachka2021,zhang2020}, studies have demonstrated that the population trends based on the complete checklists within the eBird database closely match those found in governmental monitoring surveys provided that a sufficient numbers of checklists were taken and reported within the same region \citep{horns2018,stuber2022}.

Within this article, we develop a model for data from the eBird database that is designed not only to reflect, but to help us better understand and identify the migratory trends present within this dataset. As part of this effort, we downloaded all completed checklists from the eBird database submitted during the years 2013 to 2017 within the seventeen easternmost U.S. states and Washington D.C. (a total of over five million checklists representing over five million combined hours of bird watching). This geographic region was chosen as it contains all U.S. states generally considered to be part of the Atlantic migratory flyway \citep{fritts2022}. Data were aggregated temporally by week and spatially by county, for a total of 260 weeks and 805 counties (or county-equivalents). For each week-county pairing we obtain the vector containing total counts of each species observed, as well as the total combined hours of bird watching represented by all checklists submitted within each week-county pairing. The observation efforts that produced these data are highly variable in space and time, with some counties (often those containing major population centers or state/national parks) having hundreds of hours of combined observations during a given week, while many pairings have no observations whatsoever. Approximately 29\% of the week and county combinations are represented by no checklists, indicating the potential ecological value in a model which can predict species trends for regions and time periods with little to no observational data.

\begin{figure}[!t]
    \centering
    \includegraphics[width = .9\textwidth]{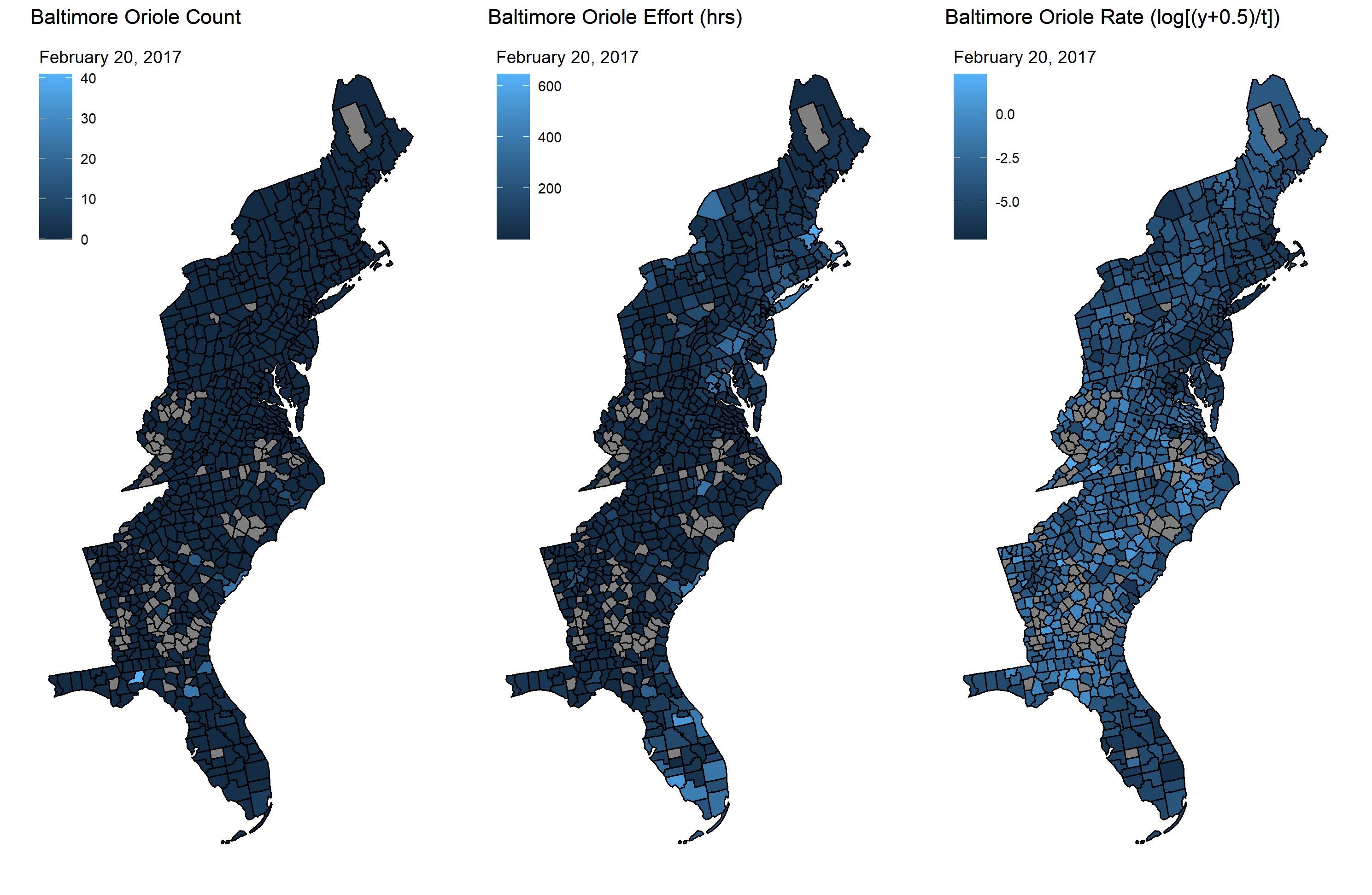}
    \caption[Baltimore oriole data from week beginning February 20, 2017]{Counts, efforts, and log observation rates for Baltimore orioles from week beginning February 20, 2017. Grey regions indicate counties with zero submitted checklists.}   \label{oriole}
\end{figure}

\begin{figure}[!t]
    \centering
    \includegraphics[width = .95\textwidth]{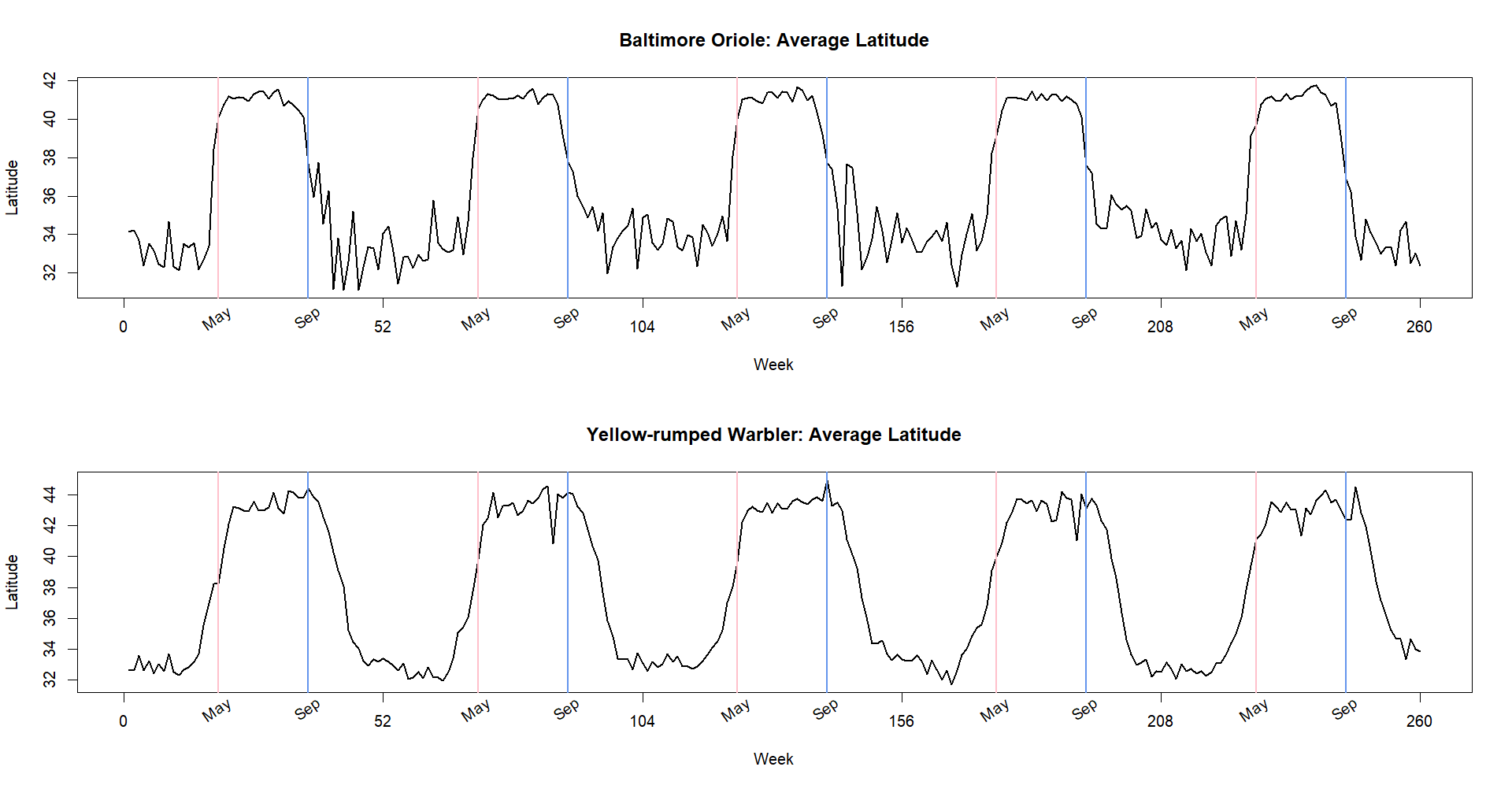}
    \caption[``Average latitude" of migratory species over time]{``Average latitude" of the Baltimore oriole and yellow-rumped warbler over time. County latitudes averaged, weights given by county-level counts per hour. The average location moves north by several degrees every May and south every September.}\label{avglat}
\end{figure}

Figure \ref{oriole} contains a visualization for observations of the Baltimore oriole, a migratory bird common to the eastern United States, during the week beginning on February 20, 2017. It depicts the county-level counts of Baltimore orioles, observation efforts, and the log rates of observed birds per hour. Figure \ref{avglat} provides a visualization of the ``average latitude" within the eastern United States over time for the Baltimore oriole and the yellow-rumped warbler, another migratory species found in this region, computed using the weighted average latitude of all county centroids, with the count of birds observed per hour for each county as weights. Average latitude increases every spring and decreases every fall illustrating a strong cyclical pattern of migration for this species, and demonstrates that the empirical trends within this data are consistent with our expectations regarding migratory birds.

Within the following section we establish some of the notation, background, and conceptual motivation for our model for observations of migratory species found within the eBird dataset. This includes a discussion of precedent for the use of circuit theory in ecological applications. In section 3 we present the full model and discuss its construction, parameterization and interpretation. We also provide details regarding prior specification and computational implementation, along with it's performance using simulated datasets. In section 4 we provide results from our analysis of the eBird data before concluding with a discussion of ways the model may be used to better understand certain migratory trends, along with remarks on potential model extensions and future work we would like to undertake as part of this research.



\section{Background}


To model the observed species counts within the eBird dataset, we define $\mathcal{D}$ as the spatial domain consisting of the seventeen US states and the District of Columbia as depicted in Figure \ref{oriole}. The 805 counties contained within $\mathcal{D}$ form a partition that may be naturally represented using the graphical structure $G = (V,E)$ where $V$ is a set of nodes corresponding to each county within $\mathcal{D}$ and $E$ is the set of edges corresponding to pairs of adjacent counties. The structure of $G$ may be mathematically represented by $A$, the $805 \times 805$ adjacency matrix constructed such that $a_{jj} = 0$, and $a_{jk} = 1$ if county $j$ is adjacent to county $k$ (denoted $j \sim k$) and $a_{jk} = 0$ otherwise. As noted in the previous section, species counts were aggregated spatially by county and temporally by week. For a single species, e.g. the Baltimore oriole, let $\by_i$ be the length-805 vector containing the total observed specimen counts during week $i \in \{1,...,260\}$ for each county in $\mathcal{D}$. We define $\bt_i$ to be the accompanying vector of observation efforts, containing the combined hours of bird watching within each county during week $i$. If no complete checklists were submitted for county $j$ during week $i$ then $t_{ij}=y_{ij}=0$. We propose modeling $\{\by_1,..,\by_{260}\}$ using a hidden Markov model that incorporates a novel temporally evolving spatial transition structure designed to characterize migratory trends over time. This class of dynamic model is commonly used in the study of spatiotemporal and network data, and can be generalized to settings with non-Gaussian data, nonlinear dependence, and non-additive noise \citep{wikle2010}. 

\subsection{Hidden Markov models}

Suppose we are given a collection of random vectors $\{\by_i\}_{1:n}$ which are observed at discrete times $i \in \{1,...,n\}$ within a spatial domain $\mathcal{D}$. Let $\by_i = \left(y(\bs_1;i),...,y(\bs_p;i)\right)'$, where $y(\bs_j;i)$ is a noisy observation of some latent spatio-temporal process $\{z(\bs;i)\}_{\bs \in \mathcal{D}}$ at time $i$ and location $\bs_j \in \mathcal{D}$. We wish to model $\bz_i$, the latent process at time $i$ realized at the same set of $p$ locations $\{\bs_1,...,\bs_p\} \in \mathcal{D}$ where $\by_i$ was observed. A first-order hidden Markov model (HMM), utilizes the following conditional independence assumptions, where $\pi(\by_i|\bz_i)$ denotes the distribution of $\by_i$ conditional on $\bz_i$:
\begin{equation}
\begin{aligned}
    &\pi(\by_i|\bz_1,...,\bz_n) = \pi(\by_i|\bz_i) \\
    &\pi(\bz_i|\bz_{i-1},...,\bz_1,\bz_0) = \pi(\bz_i|\bz_{i-1}).
\end{aligned}
\end{equation}
This implies that the joint distribution of $(\by_1,...,\by_n,\bz_1,...,\bz_n)$ may be written as follows:

\begin{equation}
    \pi(\by_1,...,\by_n,\bz_1,...,\bz_n) = \Pi_{i=1}^n [\pi(\by_i|\bz_i)\pi(\bz_i|\bz_{i-1})] \pi(\bz_0)
\end{equation}

HMMs have seen widespread usage in many environmental and ecological applications \citep{mcclintock2020,glennie2023,holsclaw2016} especially those involving models for animal movement \citep{hooten2017,patterson2008}. As shown in \cite{thygesen2009}, which models the location of fish populations using a HMM, this model may be thought of as the discrete (in space and time) approximation of the advection-diffusion equation which describes the transfer of particles or energy within a physical system and which has been used to characterize models for animal movement \citep{turchin1998,prima2018}.

Hidden Markov models are generally constructed using two components, the first being a model for observed data $\by_i$ given the latent process $\bz_i$, while the second component characterizes the evolution of the latent process in time. The most commonly used model for $\bz_i|\bz_{i-1}$ is to specify that $\mathbb{E} \left[\bz_i\right] = \bM\bz_{i-1}$, where $\bM$ is a $p \times p$ matrix of transition probabilities or weights characterizing between-location dependence \citep{wikle2010,mcclintock2020}. 

While there are many ways to define the transition matrix $\bM$, a common choice when dealing with spatial or graphical data is to define transition probabilities as a one-step random walk, such that $m_{jk} > 0$ if $j \sim k$ or $j = k$ and $m_{jk}=0$ otherwise \citep{thygesen2009,mcclintock2012}. While random walks are commonly used to model the movement of individual animals \citep{turchin1998,fagan2014,ahmed2023}, there are two reasons such a parameterization for $\bM$ is inappropriate for this application. Firstly, our interest is in modeling the underlying distribution of a migratory species over time rather than the movement of individual birds. Indeed the crowdsourced nature of the eBird data precludes the possibility of linking any two observations to the same specimen, a contrast to telemetry data where all positions of a tagged individual over time are known. (See \citet{turchin1991} for early work linking models for the movement of individuals to models for the evolution of a population's distribution over time.) Secondly, and of greater practical importance for our application, consider the significance of timescale when evaluating the possible behavior of a discrete time random walk. If $\bM$ is constructed such that $m_{jk}=0$ for all $j \nsim k$, then it will be impossible for the overall spatial distribution of $\bz_i$ to shift more than one step in any direction from its state at time $i-1$. Given that our data has a weekly temporal resolution, and the fact that some species of bird may cover distances of hundreds of kilometers per day \citep{hedenstrom1998}, it will be necessary to define a transition structure that allows for communication between non-adjacent nodes within a single time step. This can also be seen in Figure \ref{avglat}, which illustrates that significant shifts in the empirical spatial distribution of the data can occur within a single time step, especially during periods of high migration.

\subsection{Circuit theory and ecology}
The hidden Markov model and parameterization for $\bM$ presented in this article were developed in part using ideas from the fields of electrical physics and circuit theory. The introduction of electrical circuit theory to ecological models for animal movement and landscape connectivity dates to work published by Brad McRae and collaborators \citep{mcrae2006,mcrae2007,mcrae2008}. Such methods, which use the physical and mathematical laws governing voltage, current, and resistance to represent animal movement and other ecological processes have been demonstrated to perform well in many applications, and are now widespread in ecology and (increasingly) statistics \citep{dickson2019, peterson2019}. For instance, electrical current may be used as a conceptual proxy for the movement of animals through space \citep{koen2014,grafius2017}, and resistance distance---a metric intrinsically related to the notion of random walks on graphs \citep{chandra1996}---is useful for characterizing proximity between locations as a function of flow within an ecological network, especially in instances in which ecological barriers such as mountain ranges or rivers may inhibit animal movement \citep{hanks2013,thiele2018}. \citet{christensen2024a} defined a graphical covariance model using a metric closely related to resistance distance and demonstrated how it could be used to understand interactions between species distribution and environment for bird populations in North Carolina.

\begin{figure}[!t]
    \centering
    \includegraphics[width = .6\textwidth]{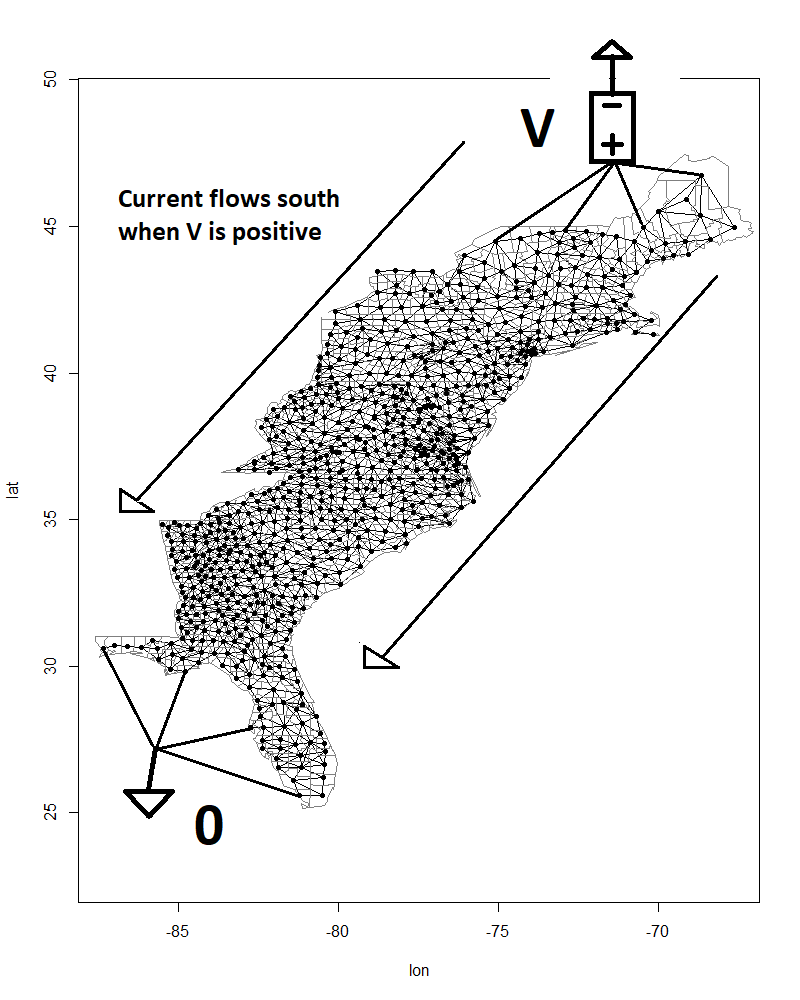}
    \caption[US counties visualized as electrical network]{US counties as a circuit with power source attached to northern counties and grounded at southern counties.}\label{circuit}
\end{figure}

To introduce the model used within this article, we consider the following analogy, illustrated by Figure \ref{circuit}. Imagine the counties of the United States as nodes within an electrical network, with resistors placed between each pair of adjacent counties. If a power source of voltage $v$ and a ground (which by definition has voltage 0) were attached to two locations within the network, electrical current would be induced between all pairs of nodes within the network, the direction and magnitude of which would be determined by the voltage $v$ and within network locations of the power source and ground. If $v$ is positive, current will flow from the power source and through the network in the direction of the ground. If $v$ is negative, current will flow in the opposite direction towards the power source. Furthermore, if $v$ and the resistances between each pair of adjacent nodes are known, it is possible to use Kirchoff's and Ohm's circuit laws (see \citet{hankin2006}) to compute the effective current, resistance and voltage differential between every pair of nodes in the network. We demonstrate how this is done in the following subsection.

\subsection{Obtaining effective resistances and currents}

We begin with $\bA$, the $805 \times 805$ unweighted adjacency matrix representing the neighborhood structure of all counties within our spatial domain. Under the previous analogy, $\bA$ represents resistances of all edges within the electrical network. We augment our graph by defining two new nodes, one (hereafter the battery node) which is adjacent to 23 counties along the northern border of our spatial domain, and a second (the ground node) adjacent to 23 counties along the interior coast of Florida. Let $\bA^*$ be the adjacency matrix for the augmented graph. We can construct $\bL = \text{diag}(\bA^* \bone_p) - \bA^*$, the Laplacian matrix of the graph. The resistance distance $\Omega_{jk}$ between nodes $j$ and $k$ is calculated as follows:

\begin{equation}\label{omega}
    \Omega_{jk} = (\be_j - \be_k)'\bL^+(\be_j-\be_k)
\end{equation}
where $\bL^+$ is the Moore-Penrose generalized inverse of $\bL$ and $\{\be_1,...,\be_{807}\}$ are the standard basis vectors. This metric characterizes the effective resistance between pairs of nodes in resistor networks and has seen widespread use within ecological applications, as previously noted \citep{klein1993, dickson2019}.

Suppose the voltage associated with the battery node is fixed to 1. Let $\bv$ be the voltages associated with all nodes within our network. By definition, the voltage at the ground node is 0, and all other nodes will have voltages between 0 and 1. Let $\bi$ be the vector of net current into each node. By Kirchoff's first law, this value will be 0 for all nodes except for the battery and ground. Ohm's law may be stated as 
\begin{equation}\label{ohm}
    \bL\bv = \bi
\end{equation}
where $\bL$ is the Laplacian matrix as before \citep{paul2001}. The only unknowns within this system are the currents at the battery and ground nodes, and the voltages at all other nodes. \citet{hankin2006} shows that this results in the following partitioned matrix equation:

\begin{equation}\label{partition}
    \begin{pmatrix}
        \bL_{11} & \bL_{12} \\
        \bL_{21} & \bL_{22}      
    \end{pmatrix} 
    \begin{pmatrix}
        \bv^{k}  \\
        \bv^{u}      
    \end{pmatrix} = 
        \begin{pmatrix}
        \bi^{u}  \\
        \bi^{k}      
    \end{pmatrix}
\end{equation}
where the superscripts $k$ and $u$ correspond to the known and unknown components of $\bv$ and $\bi$, while $\bL_{11}$ is the $2 \times 2$ portion of the Laplacian corresponding to the indices of the battery and ground nodes. The solution to this system is:
\begin{equation}\label{system}
    \begin{aligned}
        \bv^u &= \bL_{22}^+\left(\bi^k - \bL_{21}\bv^k\right) \\
        \bi^u &= \left(\bL_{11} - \bL_{12}\bL_{22}^+\bL_{21}\right)\bv^k + \bL_{12}\bL_{22}^+\bi^k.
    \end{aligned}
\end{equation}

Once the full vector $\bv$ has been obtained, the currents matrix $\bC$ is defined as follows:
\begin{equation}\label{curmat}
         c_{jk} = (v_k - v_j)/\Omega_{jk} \text{ for } j \neq k, \text{ else }c_{jk}=0.
\end{equation}

Here $c_{jk}$ represents the effective current from county $k$ towards county $j$. From Equation \ref{curmat} one can see that $c_{jk} = -c_{kj}$, and thus $\bC= -\bC^\top$. If $\bC$ is computed using a battery node voltage of 1, it can be shown that the currents matrix when the battery node has voltage $v^*$ is equal to $v^*\bC$. This in conjunction with the skew-symmetry of $\bC$ means that $\bC^\top$ is the currents matrix when the battery node has voltage equal to $-1$. 

Our general modelling approach is to define a hidden Markov model with temporally varying transition structure that can adapt to the shifting patterns of migratory flow throughout the year. The transition structure is conceptually based on how currents within the network shown in Figure \ref{circuit} (with power source in the north and ground in the south) would adapt in response to voltage changing over time. For instance, the northward spring migration may correspond to a period of negative voltage, causing currents to flow towards the power source, while the southward fall migration would be produced by positive voltage. Within the following section we formalize our mathematical notation for the model and discuss specific choices made regarding its parameterization.



\section{Method}
\subsection{Proposed model}
We analyze five years worth of completed checklists submitted to the eBird database and reported within $\mathcal{D}$, the spatial domain consisting of 805 counties within the eastern United States and depicted in Figure \ref{oriole}. Let $\bC$ be the $805 \times 805$ matrix of effective currents between all counties as defined by Equation \ref{curmat}, and let $\bD$ be the $805 \times 805$ matrix of Euclidean distances between county centroids. For a given bird species (e.g. the Baltimore oriole) let $y_{ij}$ be the combined count of specimens observed by all bird watchers during week $i \in \{1,...,260\}$ within county $j \in \{1,...,805\}$, with $\by_i$ the vector of counts across all counties during week $i$. Let $t_{ij}$ be the combined number of bird watching hours for that week-county pairing. We define the latent variable $z_{ij}$ as the expected number of birds (of the given species) one would observe during a single hour of bird watching. A straightforward modelling choice for $\by_i|\bz_i,\bt_i$ is to use a Poisson distribution:

\begin{equation}
    y_{ij} \sim \text{Poisson}(z_{ij}t_{ij}), \quad z_{ij} > 0.
\end{equation}
In addition to the standard HMM conditional independence assumption that $\pi(\by_i|\bz_1,...,\bz_{260}) = \pi(\by_i|\bz_i)$, we also assume that $\pi(y_{ij}|\bz_i) = \pi(y_{ij}|z_{ij})$.  Latent vectors $\{\bz_i\}_{1:260}$ represent the rate at which birds are observed (or would have been observed for week-county combinations with no data), but can be interpreted as reflecting relative species abundance across our spatio-temporal domain. This interpretation requires us to make the assumption that the underlying rate at which birds are observed is purely a function of abundance and not of factors such as time of observation, remoteness of location, or expertise of the individual observer. While these assumptions are almost certainly violated at the level of individually submitted checklists, on aggregate much of this bias may be mitigated, as suggested by \citet{horns2018}. Variable observation effort is accounted for by including $t_{ij}$ within the expectation of $y_{ij}$. While sampling efforts are uneven in both space and time (a greater number of checklists are reported during spring and summer months and close to major population centers) we assume that $\bt_i \perp \bz_i | \bz_{i-1},\bz_{i+1}$, meaning that there are no systematic differences in underlying species abundance between high- and low-observation regions.

To characterize how the latent process evolves in time, we specify $\bz_i|\bz_{i-1}$ as follows:

\begin{equation}
\begin{aligned}
    \bz_i &= \bM(\btheta_i)\bz_{i-1} \odot \bepsilon_i \\
    \epsilon_{ij} &\overset{iid}{\sim} \text{Gamma}(\alpha,\alpha), \quad \alpha > 0
\end{aligned}
\end{equation}
Here $\bepsilon_i$ is a vector of multiplicative, Gamma-distributed noise, and $\bM_i = \bM(\btheta_i)$ is a $805 \times 805$ transition matrix which is a function of time-indexed parameters $\btheta_i$ (greater detail is provided later in this section), and $\alpha$ is a parameter controlling the concentration of $z_{ij}$ about its mean. Our model for $z_{ij}$ may equivalently be written as:
\begin{equation}
    z_{ij} \sim \text{Gamma}(\alpha, \alpha/[\bem_{ij}'\bz_{i-1}]), \quad \alpha > 0
\end{equation}
where $\bem_{ij}$ is the $j^\text{th}$ row of $\bM_i$, meaning that $\mathbb{E}[z_{ij}] = \bem_{ij}'\bz_{i-1}$ and $\text{Var}[z_{ij}] = \bem_{ij}'\bz_{i-1}/\alpha$.

\subsection{Transition structure}

To ensure that all elements of $\bM_i$ and $\{\bz_i\}_{1:260}$ are greater than 0, we will use only positive currents when defining transition probabilities. Let $\bC^+ = \bC \vee \bzero$, setting all negative elements of $\bC$ to zero, be the currents matrix characterizing (southward) network flow when voltage is positive. Likewise, let $\bC^- = \bC^\top \vee \bzero$ be the currents matrix characterizing (northward) network flow when voltage is negative. Figure \ref{cur_ny} depicts the effective currents from Albany County, New York (in grey) to all other counties based on the corresponding column of $\bC^+$. One notes that the effective currents to all counties north of Albany County are equal to zero, and that currents to the southernmost counties in the network are largest. This latter feature indicates that raw currents alone can't be used to construct $\bM_i$, as every transition would push the majority of the mass in $\bz_{i-1}$ to one of the network poles. To account for the this, we incorporate the physical distances between locations (as contained in $\bD$) in conjunction with the currents matrices $\bC^+$ and $\bC^-$. We parameterize $\bM_i$ as follows:

\begin{figure}[!t]
    \centering
    \includegraphics[width = .3\textwidth]{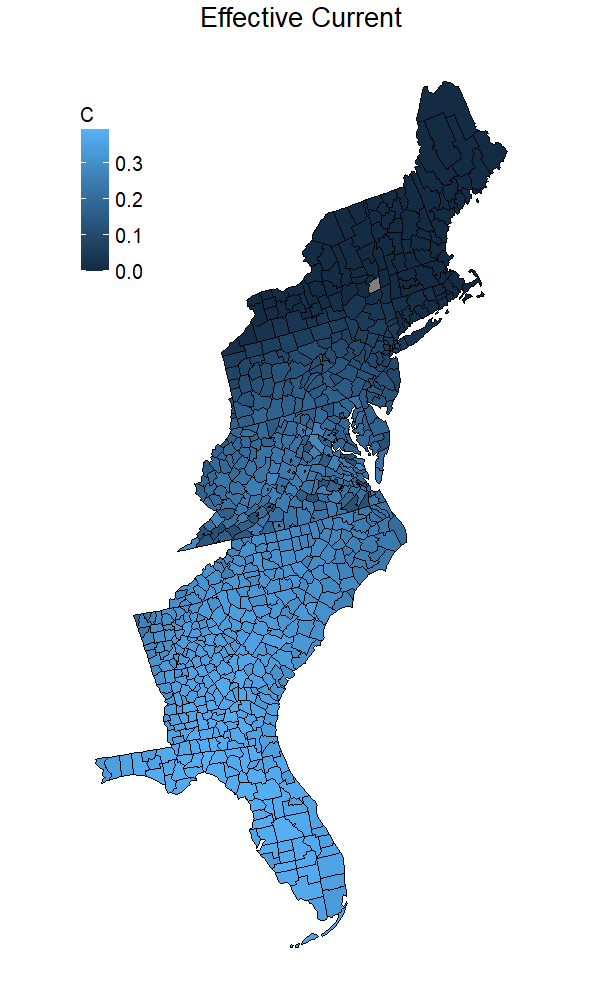}
    \caption[Effective current out of Albany County]{Effective current from Albany County, NY (in grey) towards all other counties when battery node voltage is 1.}\label{cur_ny}
\end{figure}

\begin{equation}
\begin{aligned}
    \bM(q_i,\rho_i,\nu_i,\delta_i) = \nu_i \text{ scaleCol} \left( \bC^{+/0/-} \oslash \bD^{\circ \rho_i} \right) + \delta_i\bI, \quad q_i \in \{-1,0,1\}, \; \rho_i,\nu_i,\delta_i > 0.
\end{aligned}
\end{equation}

Here $q_i \in \{-1,0,1\}$ is a parameter determining the direction of current in our network. If $q_i = 1$, then $\bC^{+/0/-} = \bC^+$. If $q_i = -1$, then $\bC^{+/0/-} = \bC^-$. If $q_i = 0$, then $\bC^{+/0/-} = \bC^0 = \bone \bone^\top$ and the system is in a state of diffusion rather than flow in a particular direction. The function $\oslash$ indicates element-wise division while defining $0/0 = 0$, and $\bD^{\circ \rho_i}$ is the matrix of between-county distances raised element wise to the $\rho_i$ power. The purpose of $\rho_i$ is to mitigate the effect seen in Figure \ref{cur_ny} in which current is strongest towards the poles of the network. Constructing the transition matrix using only $\bC$ would be akin to assuming that the majority of migrating birds always wind up in either Florida or Maine after one week, hence division by $\bD^{\circ \rho_i}$. When $\rho_i$ is large, shifts from $\bz_{i-1}$ to $\bz_i$ will cover smaller distances, while small $\rho_i$ can lead to dramatic shifts in latent spatial distribution. Because changing $\rho_i$ can significantly affect the scale of $\bC^{+/0/-} \oslash \bD^{\circ \rho_i}$ we incorporate the function $\text{scaleCol}(\cdot)$, which takes a matrix as input and rescales the columns such that they all sum to one. Lastly, $\nu_i$ and $\delta_i$ concern the rate of flow vs. self-transmission. If $\nu_i$ is high, the behavior dictated by $\bC^{+/0/-} \oslash_0 \bD^{\circ \rho_i}$ will dominate the transition pattern at time $i$. If $\delta_i$ is high, then $\bz_i$ will tend to look more like $\bz_{i-1}$. If overall species abundance within $\mathcal{D}$ is constant, $\nu_i+\delta_i$ should be equal to or close to 1. However, if there are a significant number of birds introduced to or removed from the system (via new births, deaths, or entry into or exit from Canada and Mexico) it is possible that this sum will deviate from that baseline.

\begin{figure}[!t]
    \centering
    \includegraphics[width = .85\textwidth]{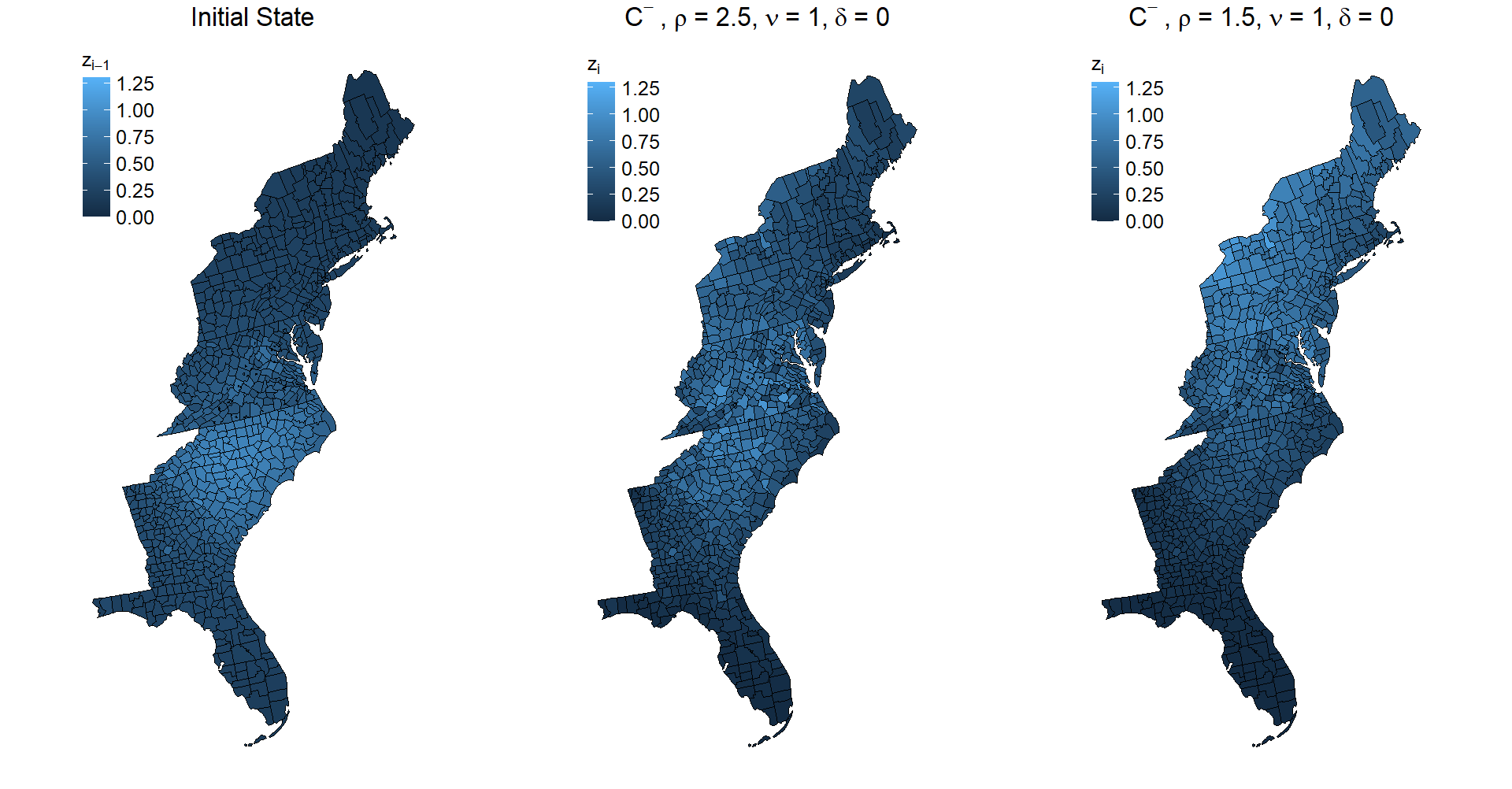}
    \caption[Impact of $\rho_i$ on transition structure]{Left: Initial state of $\bz_{i-1}$ for this figure, and Figures \ref{charge_ill} and \ref{vd_ill}. Center: $\bz_i$ with $q_i = -1$ (northward flow) and larger $\rho_i$. Right: $q_i = -1$, smaller $\rho_i$ leading to greater northward flow.}\label{rho_ill}
\end{figure}

\begin{figure}[!t]
    \centering
    \includegraphics[width = .85\textwidth]{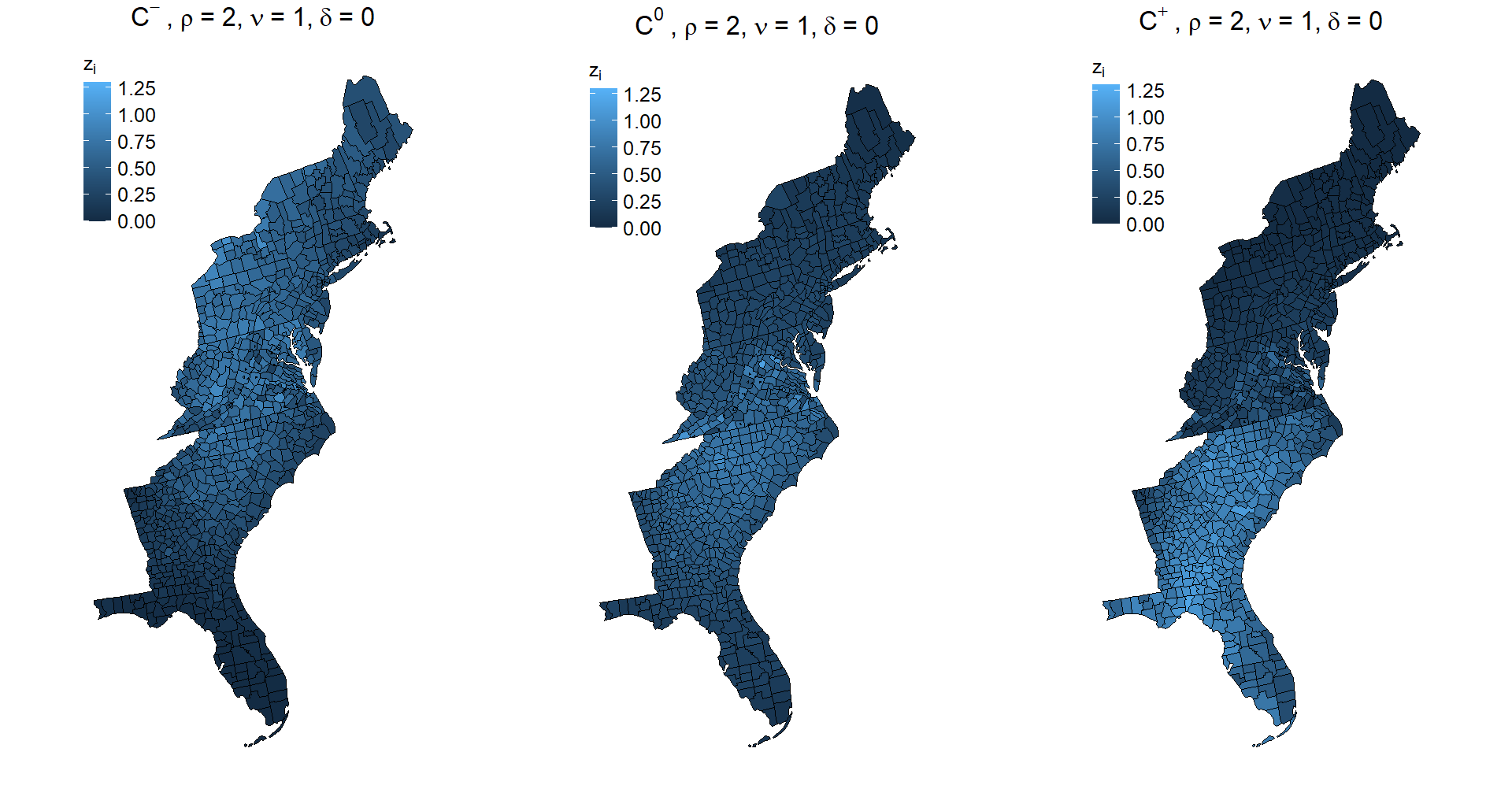}
    \caption[Impact of $q_i$ on transition structure]{Left: $\bz_i$ with $q_i = -1$ (northward flow). Center: $q_i = 0$ (diffusion). Right: $q_i = 1$ (southward flow).}\label{charge_ill}
\end{figure}

\begin{figure}[!t]
    \centering
    \includegraphics[width = .85\textwidth]{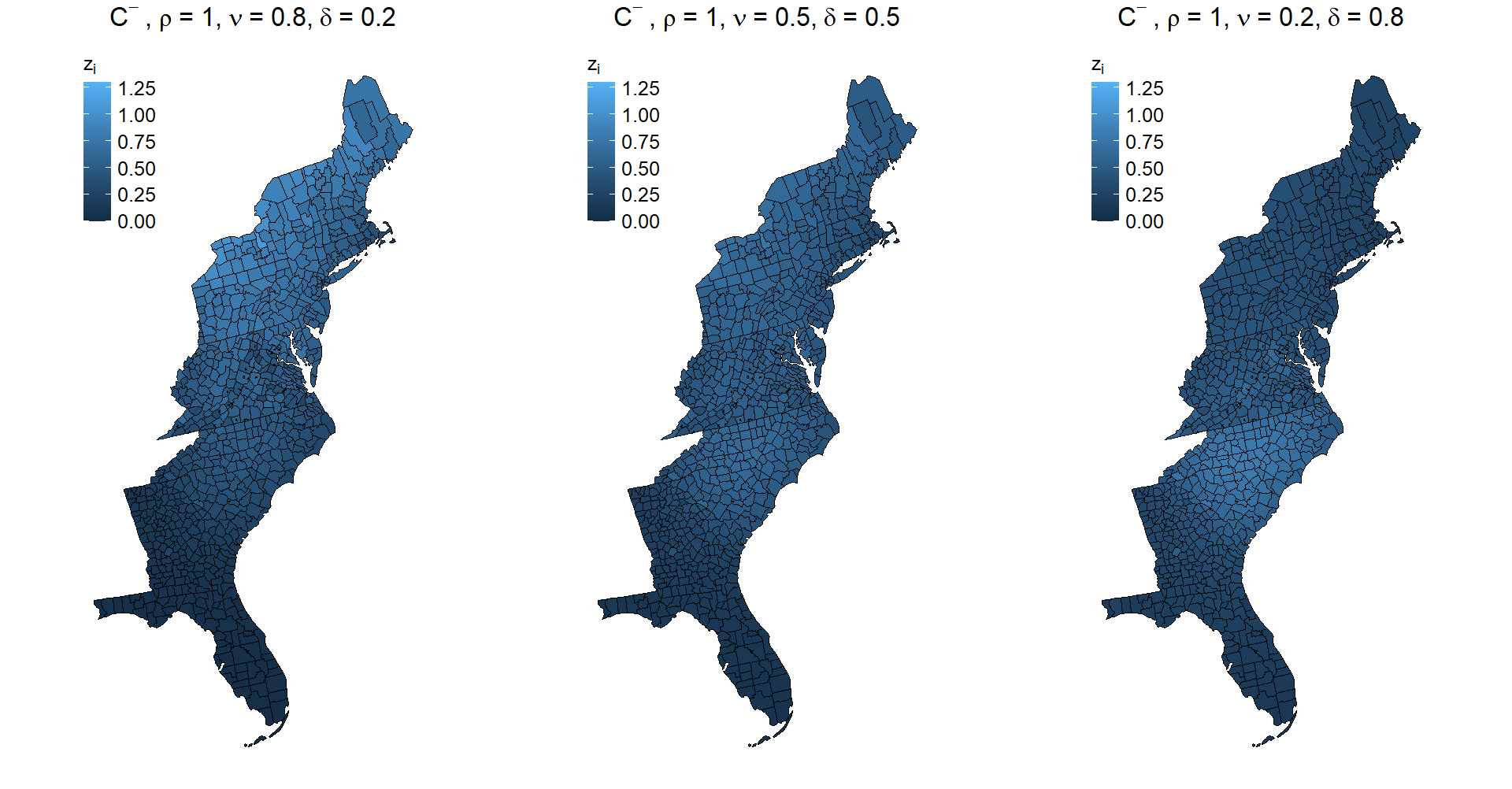}
    \caption[Impact of $\delta_i$ and $\nu_i$ on transition structure]{Left: $\bz_i$ with large $\nu_i$ relative to $\delta_i$. Center: Equal $\nu_i$ and $\delta_i$. Right: Small $\nu_i$ relative to $\delta_i$.}\label{vd_ill}
\end{figure}

\begin{figure}[!t]
    \centering
    \includegraphics[width = .85\textwidth]{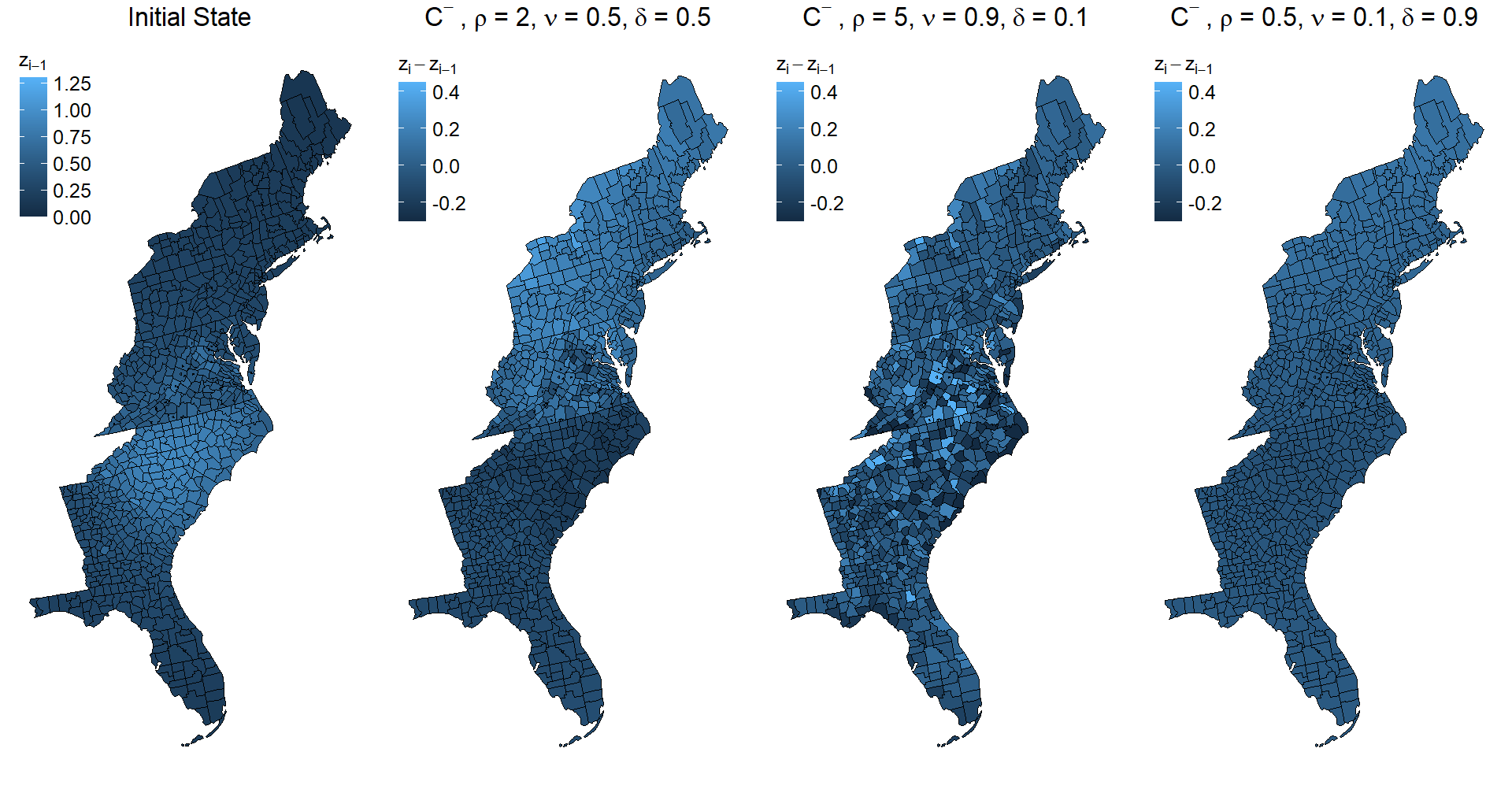}
    \caption[Additional visualization for transition matrix parameters]{Change in latent rate $(\bz_i-\bz_{i-1})$ under different parameter settings.}\label{contrast}
\end{figure}

To illustrate the behavior of our model, consider Figures \ref{rho_ill}, \ref{charge_ill}, \ref{vd_ill} and \ref{contrast}, which depict the transitions between latent states under different parameter settings. The leftmost subfigure in Figure \ref{rho_ill} represents the initial state of $\bz_{i-1}$ while the other eight subfigures from \ref{rho_ill}, \ref{charge_ill} and \ref{vd_ill} represent the state of $\bz_i$ after transition according to the indicated parameters. As seen in Figure \ref{rho_ill}, the initial state is characterized by high concentration in the Carolinas and relatively low concentration in all other regions. Figure \ref{rho_ill} depicts the impact of $\rho_i$ on the transition from $\bz_{i-1}$ to $\bz_i$. The rightmost subfigure, with a smaller value of $\rho_i$ depicts greater concentration of mass in the northern portions of the spatial domain, whereas higher $\rho_i$ leads to a northward shift that is more muted. Figure \ref{charge_ill} depicts the influence of $q_i$. The left subfigure depicts northward flow, the central subfigure diffusion, and the right southward flow, with all other parameters constant. Figure \ref{vd_ill} depicts the impact of $\nu_i$ and $\delta_i$. While all three subfigures  here depict northward flow under the same value of $\rho_i$, the rightmost figure differs notably less from the initial state due to higher $\delta_i$. Lastly, consider the illustration in Figure \ref{contrast}, depicting the change from $\bz_{i-1}$ to $\bz_{i}$ under different parameter settings. Note that while all are characterized by northward flow, the transition structure characterized by moderate $\rho_i$ and balanced $\nu_i$ and $\delta_i$ (second subfigure) produces a more substantive northward shift in latent rates than the transition matrix produced by high $\rho_i$ and high $\nu_i/(\nu_i+\delta_i)$ (third subfigure), and the matrix produced by low $\rho_i$ and low $\nu_i/(\nu_i+\delta_i)$ (fourth subfigure). This indicates that each transition matrix parameter can take on values that stifle flow, even if all other parameters are valued such that flow would be encouraged.

\subsection{Priors and model fitting}

We perform inference for the parameters of this model by approximating the posterior distribution under the following prior distribution:

\begin{equation}\label{prior3}
\begin{aligned}
    \alpha &\sim \text{Gamma}(a_\alpha,b_\alpha) \\
    q_i &\overset{iid}{\sim} \begin{cases}
    \begin{aligned}
    &P(q_i = 1) \;= p_+\\[-1ex]
    &P(q_i = 0) \;= p_0\\[-1ex]
    &P(q_i = -1) = p_-\\
    \end{aligned}
    \end{cases} \\
    \rho_i &\overset{iid}{\sim} \text{Gamma}(a_\rho,b_\rho) \\
    \nu_i &\overset{iid}{\sim} \text{Gamma}(a_\nu,b_\nu) \\
    \delta_i &\overset{iid}{\sim} \text{Gamma}(a_\delta,b_\delta).
\end{aligned}
\end{equation}
Prior parameters and initial latent state vector $\bz_0$ should be chosen in an application appropriate manner. The posterior distribution $\pi\left(\{\bz_i\}_{1:260},\{q_i\}_{1:260},\{\rho_i\}_{1:260},\{\nu_i\}_{1:260},\{\delta_i\}_{1:260},\alpha|\{\by_i\}_{1:260}\right)$ is obtained using the Markov chain Monte Carlo (MCMC) sampler described in Algorithm \ref{alg}. Greater detail regarding the sampler and proposal distributions used is provided in Appendix C.

\begin{algorithm}[!ht]
\caption{MCMC Overview}\label{alg}
\begin{algorithmic}
\small
\State \textbf{Input:} Observations $\{\by_i\}_{1:n}$, $\{\bt_i\}_{1:n}$ and pre-computed currents matrix $\bC$
\State \textbf{Output:} $T$ posterior samples for parameters $\{\bz_i\}_{1:n}, \{q_i\}_{1:n}, \{\rho_i\}_{1:n}, \{\nu_i\}_{1:n}, \{\delta_i\}_{1:n}, \text{ and } \alpha$
\State Initialize $\{\bz_i^{(0)}\}_{1:n}, \{q_i^{(0)}\}_{1:n}, \{\rho_i^{(0)}\}_{1:n}, \{\nu_i^{(0)}\}_{1:n}, \{\delta_i^{(0)}\}_{1:n}, \text{ and } \alpha^{(0)}$
\For{$t = 1 \text{ to } T$}
\For{$i = 1 \text { to } n$}

\State MH update (MALA proposal) of $\bz^{(t)}_i$ based on  $\pi(\bz_i|\by_i,\bz_{i-1}^{(t)},\bz_{i+1}^{(t-1)},\bM_i^{(t-1)},\alpha^{(t-1)})$

\State Direct update of $q_i^{(t)}$ based on $\pi(q_i|\bz_i^{(t)},\bz_{i-1}^{(t)},\rho_i^{(t)},\nu_i^{(t)},\delta_i^{(t)},\alpha^{(t-1)})$
\State MH update (aMCMC proposal) of $\theta^{(t)}_i = (\rho_i^{(t)},\nu_i^{(t)},\delta_i^{(t)})'$ based on $\pi(\theta_i|\bz_i^{(t)},\bz_{i-1}^{(t)},q_i^{(t)},\alpha^{(t-1)})$

\EndFor
\State MH update of $\alpha^{(t)}$ based on $\pi(\alpha|\{\bz_i\}_{1:n},\{\bM_i\}_{1:n})$
\EndFor
\end{algorithmic}
\end{algorithm}


\subsection{Simulation}
\begin{figure}[!t]
    \centering
    \includegraphics[width = .4\textwidth]{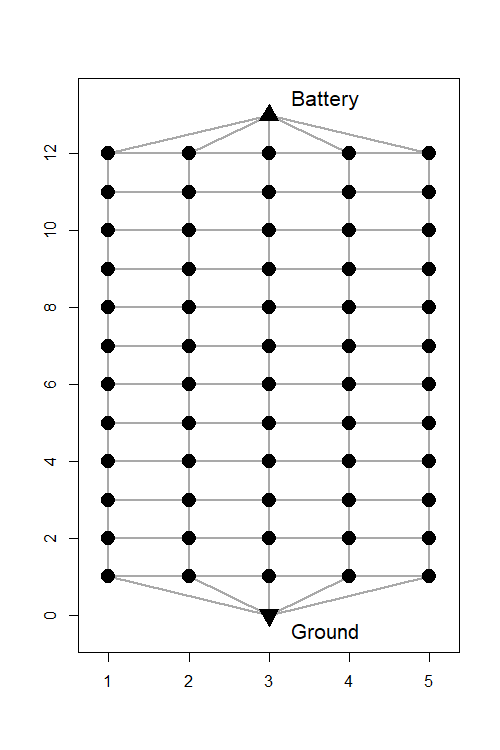}
    \caption[Network used for simulation]{Network structure used for simulation.}\label{simgrid}
\end{figure}

Before fitting our model to the full data set, we assess performance on a smaller data set by simulating data from our model on the $12 \times 5$ lattice network depicted in Figure \ref{simgrid}. Data $\{\by_i\}_{1:10}$ were produced by generating exposure efforts $t_{ij} \overset{iid}{\sim} \text{Exponential}(0.1)$ and simulating $\{\bz_i\}_{1:10}$ by initializing $\bz_0 = \bone$ and generating transition parameters $q_i \overset{iid}{\sim} \text{Multinomial}(p_+ =0.25, p_0 = 0.5, p_- =0.25)$, $\rho_i \overset{iid}{\sim} \text{Gamma}(3,1.5)$, $\nu_i \overset{iid}{\sim} \text{Gamma}(3,6)$, $\delta_i \overset{iid}{\sim} \text{Gamma}(3,6)$, and $\alpha \sim \text{Gamma}(10,2)$ for $i \in \{1,...,10\}$ time periods. In order to assess the model's ability to recover the underlying rate $z_{ij}$ in unobserved regions, 30 percent (approximately the percentage of the eBird data for which observation effort is 0) of total observations were censored at random such that $y_{ij} = t_{ij} = 0$. Samples from the posterior distribution were obtained using MCMC, from which coverage rates and other statistics were calculated. This simulation process was repeated twenty times. Tables \ref{zsim} and \ref{othersim} contain aggregated results from all model fittings. 

\begin{table}[ht]
\caption[Simulation results for recovery of latent $\bz_i$]{Average coverage, credible interval width, bias, and RMSE for $z_{ij}$ when data are observed and censored.}\label{zsim}
\centering
\begin{tabular}{rrrr}
  \hline
 & Coverage (90\%)  & Bias &  RMSE \\ 
  \hline
$y_{ij}, \; t_{ij}$ observed & 0.883  & -0.003 & 0.349 \\ 
  $y_{ij} = t_{ij} = 0$ & 0.880  & -0.002 & 0.589 \\ 
   \hline
\end{tabular}
\end{table}

\begin{table}[ht]
\caption[Simulation results for recovery of transition matrix parameters]{Average 90\% coverage rates for all transition structure parameters.}\label{othersim}
\centering
\begin{tabular}{rrrrrr}
  \hline
 & $q_i$ & $\rho_i$ & $\nu_i$ & $\delta_i$ & $\alpha$ \\ 
  \hline
Coverage (90\%) & 0.895 & 0.610 & 0.905 & 0.855 & 0.900 \\ 
   \hline
\end{tabular}
\end{table}

We note that the model recovers the underlying rate parameters $\{\bz_i\}_{1:10}$ effectively, even for nodes and time periods where no data are available, suggesting that this model could be valuable in estimating species abundance for regions with minimal observation effort. All other model parameters have appropriate coverage rates, with the exception of $\rho_i$. Investigation of simulation output indicated that increases in $\rho_i$ have diminishing impact on likelihood, making higher true values of $\rho_i$ difficult to correctly recover. In general, this issue does not seem to negatively impact inference on other parameters. We now turn to a discussion of model results when fit to observations from the eBird data set.


\section{Analysis of eBird data for migratory species}
To understand trends in the movement and relative abundance of migratory birds, we fit our model to five years worth of weekly county-level observations for the Baltimore oriole and yellow-rumped warbler. In order to reduce computational burden, the model was fit separately for each year, allowing some degree of parallelization at the cost of breaking temporal dependence between fits. Using the priors given in Equation \ref{prior3}, the following hyperparameters were chosen: $a_\rho = a_\nu = a_\delta = 5$, $b_\rho = 2$, $b_\nu= b_\delta = 10$, and $(p_+,p_0,p_-) = (0.2,0.6,0.2)$. This results in $\rho_i$ having a prior expectation of 2.5, a threshold we found to be consistent with modest but non-trivial levels of flow and diffusion when simulating current matrices, and balanced $\nu_i$ and $\delta_i$. We fixed $\alpha = 2$ based on an ad hoc empirical Bayes procedure. This decision was made in order to keep $\alpha$ consistent between the separate model fits for each year, to ensure that the the multiplicative error term has a non-zero mode, and because we found in simulation that fixing $\alpha$ significantly improves mixing times while having limited impact on inference for other parameters. Computation was performed using the Duke University Compute Cluster. An average of approximately 15000 posterior samples were obtained, requiring two weeks of time and resulting in an average effective sample size of approximately 500 after burn-in.

\begin{figure}[!t]
    \centering
    \includegraphics[width = .95\textwidth]{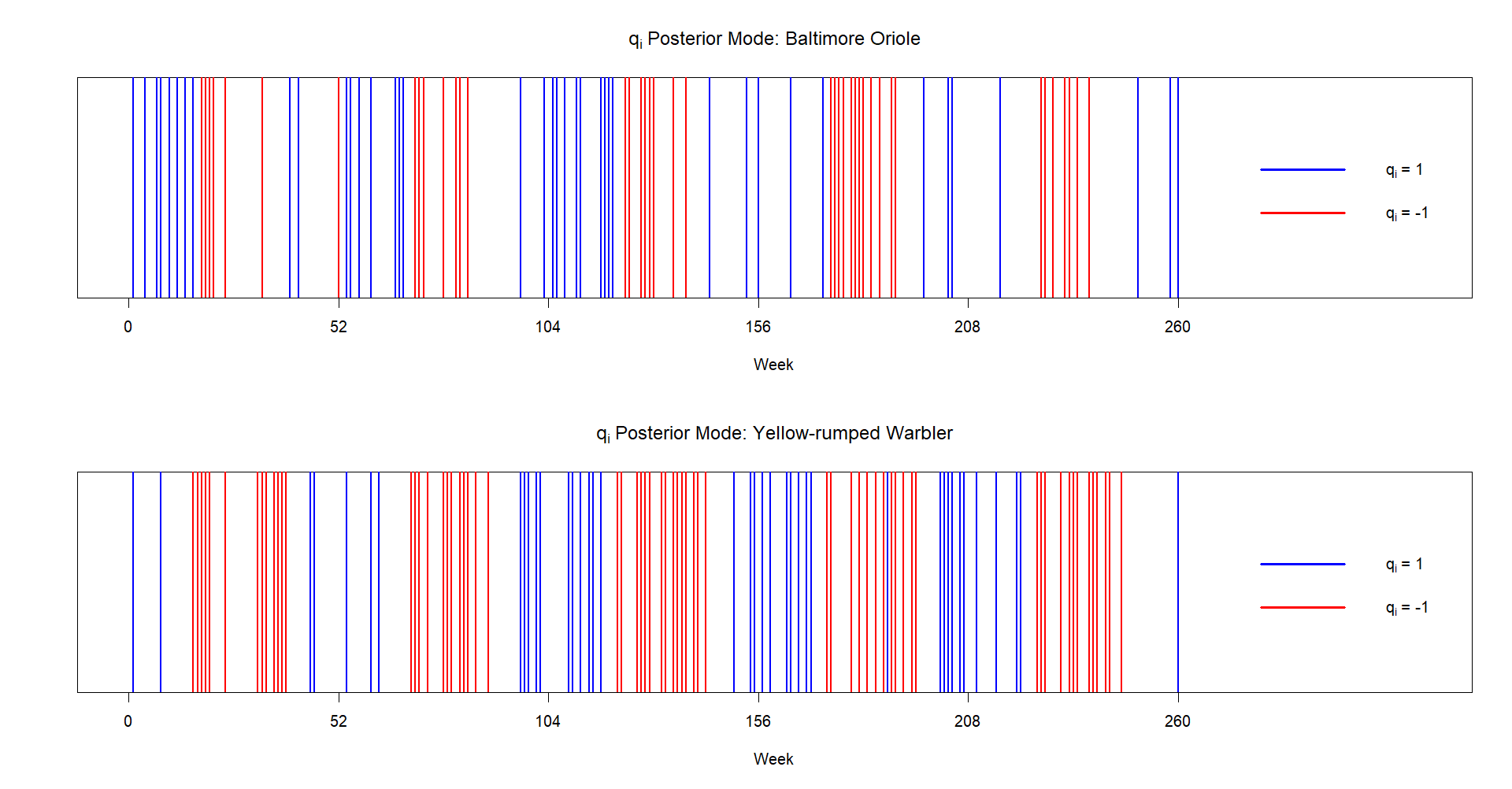}
    \caption[Posterior mode of $\bq$]{The posterior mode of $\bq$: blue lines indicate weeks with high posterior probability that current is positive, resulting in southward flow, while red lines indicate weeks with high posterior probability that current is negative resulting in northward flow.}\label{postmode}
\end{figure}

\begin{figure}[!t]
    \centering
    \includegraphics[width = .95\textwidth]{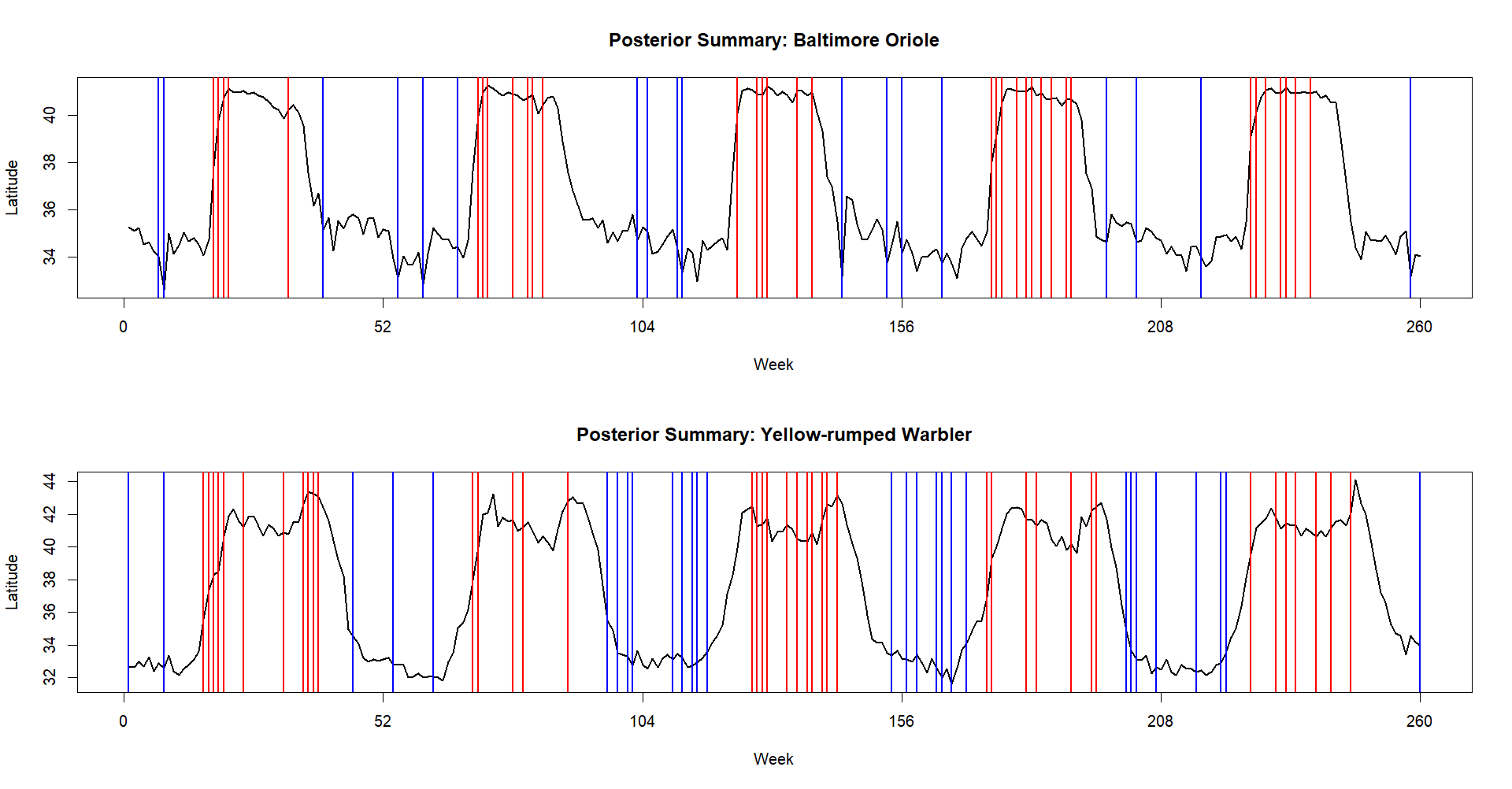}
    \caption[Migratory patterns summarized using posterior output]{Black lines depict average latitude over time based on posterior distribution of $\{\bz_i\}_{1:260}$. Vertical red lines indicate time periods of substantive northward flow and blue lines indicated periods of substantive southward flow, based on posterior of $\{q_i\}_{1:260}$, $\{\rho_i\}_{1:260}$, $\{\nu_i\}_{1:260}$, $\{\delta_i\}_{1:260}$}\label{postsummary}
\end{figure}

Within this section, our primary focus to assess how our model informs our understanding of migratory patterns over time for each species. Recall from discussion of Figures \ref{rho_ill}-\ref{contrast} that $q_i$, $\rho_i$, $\nu_i$ and $\delta_i$ all work in concert to characterize the transition behavior at time $i$. While one could simply evaluate the posterior mode of $\{q_i\}_{1:260}$ over time, depicted in Figure \ref{postmode}, and infer that southward migration is occurring when the posterior distribution of $q_i$ is concentrated at 1, and northward at $-1$, it may be useful to take other parameter estimates at time $i$ into account. For example, we may wish to characterize a period of "substantive" flow as one for which the posterior probability of $q_i \in {1,-1}$ exceeds 0.9, the posterior mean of $\rho_i > 3$ and the posterior mean of $\delta_i/(\nu_i+\delta_i) > 0.7$, thresholds corresponding approximately to the 75th percentile across $i$, thereby omitting any time periods with parameter values that significantly inhibit flow. Figure \ref{postsummary} contains plots of the average latitude over time (in black, and based on the posterior distribution of $\{\bz_i\}_{1:260}$) for the population of the Baltimore-oriole and yellow-rumped warbler, with vertical red lines indicating weeks with substantive northward $(q_i=-1)$ migration according to the above heuristic, and blue lines for southward $(q_i=1)$ migration.

As expected, our model identifies patterns of northward migration during spring and summer, with patterns of southward migration during fall and winter for each species. The consistency of certain patterns observable in Figure \ref{postsummary} is worth noting. In each of the five years analyzed, the first substantive instance of northward migration for the Baltimore oriole occurs during the 17th or 18th week of the year (corresponding to the period between April 23 to May 6 in non-leap years), while first such period for the yellow-rumped warbler occurs each year (except in 2015) during the 15th to 17th weeks. Notably, the week of initial northward migration is correlated ($R=0.72$) between these two species, suggesting potential common environmental factors driving the onset of spring migration for both species. The portion of the year during which our model is identifying northward shifts in distribution tends to last longer for the yellow-rumped warbler than for the Baltimore oriole. All of these findings are consistent with preexisting research that suggests that the yellow-rumped warbler has a more drawn out spring migratory season relative to the Baltimore oriole \citep{warbler2020,oriole2020}. 

In contrast to the patterns observed with regards to northward migration, our model identifies fewer and more temporally varied periods of southward migration for each species. Ecologists have noted differences in patterns between spring and fall migration in terms of speed and duration of travel \citep{tryjanowski2002,nilsson2013} with spring migration occurring more rapidly over a shorter period, and seasonal differences in the physiology of migrating birds \citep{sharma2018}. It has been theorized that sexual competition drives the intensity and speed of the spring migration \citep{kokko1999}, whereas fall migration is driven more by local resource availability and climate conditions \citep{nilsson2013}. Taken together, the patterns observed during spring migration tend to be more dramatic and consistent across populations, whereas fall migration is more protracted and exhibits more localized behavior, which may account for the weaker and less consistent signals of southward migration identified by our model.


\section{Discussion}
Within this article we utilized crowdsourced bird watching observations from the eBird database in order to model the latent abundance of migratory bird species over time and space. We constructed a hidden Markov model with novel transition matrix parameterization utilizing principles from circuit theory in order to reflect migratory patterns present within the data. After illustrating model behavior and demonstrating its ability to recover the true underlying distribution in locations with missing data we fit our model to observations of the Baltimore oriole and yellow-rumped warbler, identifying migratory trends which are consistent with contemporary scientific understanding.

\begin{figure}[!t]
    \centering
    \includegraphics[width = .4\textwidth]{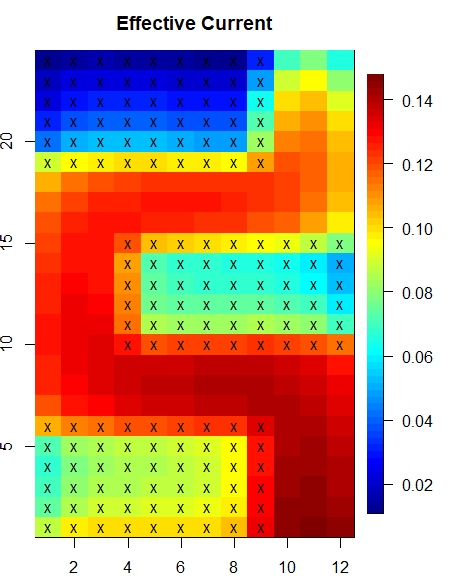}
    \caption[Impact of variable resistances on network flow]{High and low resistance regions shape flow through a network, enabling us to learn migratory pathways.}\label{var_resist}
\end{figure}

We conclude this article with a brief discussion of potential applications, model extensions, and follow up questions to the content of this article. As noted, the identification of migratory flyways (the routes birds use while travelling) is an important task in the study of avian migration \citep{buhnerkempe2016}. The model as implemented within this article did not explicitly consider this question, essentially treating the entire spatial domain as a single flyway. As we consider potentially useful future extensions to this work, a major point of interest is to explore the use of variable edge weights in the construction of our model's adjacency structure. Under the analogy of the spatial domain as electrical network, the unweighted county adjacency matrix is akin to unit resistors being used exclusively in network construction. If edge weights were instead parameterized as functions of environmental covariates or spatial random effects \citep{christensen2024b,hanks2013}, current flow could exhibit more flexible and scientifically noteworthy behaviors, especially within the context of flyway detection. As an illustration of this, see Figure \ref{var_resist}, which depicts a $24 \times 12$ lattice network with battery node attached to the top of the graph and ground at the bottom. Cells marked with ``X"s indicate high resistance regions; one could imagine that they correspond to mountains or other features inhibiting animal movement. Effective current is depicted relative to the battery node which has voltage equal to 1. As can be seen current flows around high-resistance regions and through low resistance regions. Defining a model such that edge resistances are estimated from the data would enable a detailed understanding of the paths birds are utilizing during migration; similar methods have already been used to characterize patterns of bird movement, albeit at more local scale and with resistances pre-specified according to expert belief rather than estimated \citep{grafius2017}.

Understanding the impact of climate change on bird populations and their migratory behaviors is of considerable ecological importance \citep{marra2005}. Scientists have already observed shifting patterns in the timing and duration of migration season over the past decades \citep{covino2020}. As noted in the previous section, our model identified that the onset of spring migration was correlated between the studied species. Incorporating historic climate information into our model, or comparing model output to such data could allow us to make more significant inferences about these relationships. Because climate conditions can vary dramatically across the migratory range of many species and local conditions can impact migratory timings \citep{tottrup2010}, it could be valuable to adapt our model's transition structure such that flow patterns are defined locally rather than globally.

\bibliographystyle{unsrtnat}
\bibliography{ref}

\begin{thebibliography}{63}
\providecommand{\natexlab}[1]{#1}
\providecommand{\url}[1]{\texttt{#1}}
\expandafter\ifx\csname urlstyle\endcsname\relax
  \providecommand{\doi}[1]{doi: #1}\else
  \providecommand{\doi}{doi: \begingroup \urlstyle{rm}\Url}\fi

\bibitem[Barton and Sandercock(2018)]{barton2018}
Gina~G. Barton and Brett~K. Sandercock.
\newblock Long-term changes in the seasonal timing of landbird migration on the pacific flyway.
\newblock \emph{The Condor: Ornithological Applications}, 120\penalty0 (1):\penalty0 30--46, 2018.

\bibitem[Visser et~al.(2009)Visser, Perdeck, Van~Balen, and Both]{visser2009}
Marcel~E. Visser, Albert~C. Perdeck, Johan~H. Van~Balen, and Christiaan Both.
\newblock Climate change leads to decreasing bird migration distances.
\newblock \emph{Global Change Biology}, 15\penalty0 (8):\penalty0 1859--1865, 2009.

\bibitem[Sutherland(1996)]{sutherland1996}
William~J. Sutherland.
\newblock Predicting the consequences of habitat loss for migratory populations.
\newblock \emph{Proceedings of the Royal Society of London. Series B: Biological Sciences}, 263\penalty0 (1375):\penalty0 1325--1327, 1996.

\bibitem[Buhnerkempe et~al.(1996)Buhnerkempe, Webb, Merton, Buhnerkempe, Givens, Miller, and Hoeting]{buhnerkempe2016}
Michael~G. Buhnerkempe, Colleen~T. Webb, Andrew~A. Merton, John~E. Buhnerkempe, Geof~H. Givens, Ryan~S. Miller, and Jennifer~A. Hoeting.
\newblock Identification of migratory bird flyways in north america using community detection on biological networks.
\newblock \emph{Ecological Applications}, 26\penalty0 (3):\penalty0 740--751, 1996.

\bibitem[Alerstam(1993)]{alerstam1993}
Thomas Alerstam.
\newblock \emph{Bird migration}.
\newblock Cambridge University Press, 1993.

\bibitem[Helbig(2003)]{helbig2003}
Andreas~J. Helbig.
\newblock Evolution of bird migration: A phylogenetic and biogeographic perspective.
\newblock \emph{Avian Migration}, 2003.

\bibitem[{The Cornell Lab}(2021)]{cornell2021}
{The Cornell Lab}.
\newblock The basics of bird migration: how, why, and where, 2021.
\newblock URL \url{https://www.allaboutbirds.org/news/the-basics-how-why-and-where-of-bird-migration/}.
\newblock [Online; accessed 7-April-2023].

\bibitem[Dunn and Gipson(1977)]{dunn1977}
James~E. Dunn and Phillip~S. Gipson.
\newblock Analysis of radio telemetry data in studies of home range.
\newblock \emph{Biometrics}, pages 85--101, 1977.

\bibitem[Perras and Nebel(2012)]{perras2012}
Michael Perras and Silke Nebel.
\newblock Satellite telemetry and its impact on the study of animal migration.
\newblock \emph{Nature Education Knowledge}, 3\penalty0 (12):\penalty0 4, 2012.

\bibitem[Hooten et~al.(2017)Hooten, Johnson, and McClintock]{hooten2017}
Mevin~B. Hooten, Devin~S. Johnson, and Juan~M.M. McClintock.
\newblock \emph{Animal Movement: Statistical Models for Telemetry Data}.
\newblock CRC Press, 2017.

\bibitem[Hebblewhite and Haydon(2010)]{hebblewhite2010}
Mark Hebblewhite and Daniel~T. Haydon.
\newblock Distinguishing technology from biology: a critical review of the use of gps telemetry data in ecology.
\newblock \emph{Philosophical Transactions of the Royal Society B}, 365:\penalty0 2303--2312, 2010.

\bibitem[Lindenmayer and Likens(2010)]{lindenmayer2010}
David~B. Lindenmayer and Gene~E. Likens.
\newblock The science and application of ecologoical monitoring.
\newblock \emph{Biological Conservation}, 2010.

\bibitem[Caughlan and Oakley(2001)]{caughlan2001}
Lynne Caughlan and Karen~L. Oakley.
\newblock Cost considerations for long-term ecological monitoring.
\newblock \emph{Ecological Indicators}, 1\penalty0 (2):\penalty0 123--124, 2001.

\bibitem[Schwoerer and Dawson(2022)]{schwoerer2022}
Tobias Schwoerer and Natalie~G. Dawson.
\newblock Evolution of bird migration: A phylogenetic and biogeographic perspective.
\newblock \emph{PLoS One}, 17\penalty0 (7), 2022.

\bibitem[Sullivan et~al.(2009)Sullivan, Wood, Iliff, Bonney, Fink, and Kelling]{ebird}
Brian~L. Sullivan, Christopher~L. Wood, Marshall~J. Iliff, Rick~E. Bonney, Daniel Fink, and Steve Kelling.
\newblock ebird: A citizen-based bird observation network in the biological sciences.
\newblock \emph{Biological conservation}, 142\penalty0 (10):\penalty0 2282--2292, 2009.

\bibitem[{Team eBird}(2024)]{ebird2024}
{Team eBird}.
\newblock 2023 year in review: {eBird}, {Merlin}, {Macaulay Library}, and {Birds of the World}, 2024.
\newblock URL \url{https://ebird.org/news/2023-year-in-review}.
\newblock [Online; accessed 16-April-2024].

\bibitem[Adde et~al.(2021)Adde, Casabona~i Amat, Mazerolle, Darveau, Cumming, and O'Hara]{adde2021}
Antoine Adde, Clara Casabona~i Amat, Marc~J. Mazerolle, Marcel Darveau, Steven~G. Cumming, and Robert~B. O'Hara.
\newblock Integrated modeling of waterfowl distribution in western canada using aerial survey and citizen science {(eBird)} data.
\newblock \emph{Ecosphere}, 12\penalty0 (10), 2021.

\bibitem[Tang et~al.(2021)Tang, Clark, and Gelfand]{tang2021}
Becky Tang, James~S. Clark, and Alan~E. Gelfand.
\newblock Modeling spatially biased citizen science effort through the {eBird} database.
\newblock \emph{Environmental and Ecological Statistics}, 28\penalty0 (3):\penalty0 609--630, 2021.

\bibitem[Hass et~al.(2022)Hass, La~Sorte, McCaslin, Belotti, and Horton]{haas2022}
Elaina~K. Hass, Frank~A. La~Sorte, Hanna~M. McCaslin, Maria~C.T.D. Belotti, and Kyle~G. Horton.
\newblock The correlation between {eBird} community science and weather surveillance radar-based estimates of migration phenology.
\newblock \emph{Global Ecology and Biogeography}, 31\penalty0 (11):\penalty0 2219--2230, 2022.

\bibitem[Bianchini and Tozer(2023)]{bianchini2023}
Kristin Bianchini and Douglas~C. Tozer.
\newblock Using breeding bird survey and {eBird} data to improve marsh bird monitoring abundance indices and trends.
\newblock \emph{Avian Conservation and Ecology}, 18\penalty0 (1), 2023.

\bibitem[Hochachka et~al.(2023)Hochachka, Ruiz-Gutierrez, and Johnston]{hochachka2023}
Wesley~M. Hochachka, Viviana Ruiz-Gutierrez, and Alison Johnston.
\newblock Considerations for fitting occupancy models to data from {eBird} and similar volunteer-collected data.
\newblock \emph{Ornithology}, 140\penalty0 (4), 2023.

\bibitem[Hochachka et~al.(2021)Hochachka, Alonso, Gutierrez-Exposito, Miller, and Johnston]{hochachka2021}
Wesley~M. Hochachka, Hany Alonso, Carlos Gutierrez-Exposito, Eliot Miller, and Alison Johnston.
\newblock Regional variation in the impacts of the {COVID-19} pandemic on the quantity and quality of data collected by the project {eBird}.
\newblock \emph{Biological Conservation}, 254, 2021.

\bibitem[Zhang(2020)]{zhang2020}
Guiming Zhang.
\newblock Spatial and temporal patterns in volunteer data contribution activities: A case study of {eBird}.
\newblock \emph{ISPRS International Journal of Geo-Information}, 9\penalty0 (10):\penalty0 597, 2020.

\bibitem[Horns et~al.(2018)Horns, Adler, and Sekercioglu]{horns2018}
Joshua~J. Horns, Frederick~R. Adler, and Cagan~H. Sekercioglu.
\newblock Using opportunistic citizen science data to estimate avian population trends.
\newblock \emph{Biological Conservation}, 221:\penalty0 151--159, 2018.

\bibitem[Stuber et~al.(2022)Stuber, Robinson, Bjerre, Otto, Millsap, Zimmerman, Brasher, Ringelman, Fournier, Yetter, Isola, and Viviana]{stuber2022}
Erica~F. Stuber, Orin~J. Robinson, Emily~R. Bjerre, Mark~C. Otto, Brian~A. Millsap, Guthrie~S. Zimmerman, Michael~G. Brasher, Kevin~M. Ringelman, Auriel~M.V. Fournier, Aaron Yetter, Jennifer~E. Isola, and Ruiz-Gutierrez Viviana.
\newblock The potential of semi-structured citizen science data as a supplement for conservation decision-making: Validating the performance of ebird against targeted avian monitoring efforts.
\newblock \emph{Biological Conservation}, 270, 2022.

\bibitem[Fritts(2022)]{fritts2022}
Rachel Fritts.
\newblock Avian superhighways: The four flyways of north america, 2022.
\newblock URL \url{https://abcbirds.org/blog/north-american-bird-flyways/}.
\newblock [Online; accessed 7-April-2023].

\bibitem[Wikle and Hooten(2010)]{wikle2010}
Christoper~K. Wikle and Mevin~B. Hooten.
\newblock A general science-based framework for dynamical spatio-temporal models.
\newblock \emph{Test}, 19:\penalty0 417--451, 2010.

\bibitem[McClintock et~al.(2020)McClintock, Langrock, Gimenez, Cam, Borchers, Glennie, and Patterson]{mcclintock2020}
Brett~T. McClintock, Roland Langrock, Olivier Gimenez, Emmanuelle Cam, David~L. Borchers, Richard Glennie, and Toby~A. Patterson.
\newblock Hidden markov models: pitfalls and opportunities in ecology.
\newblock \emph{Ecology Letters}, 23\penalty0 (12):\penalty0 1878--1903, 2020.

\bibitem[Glennie et~al.(2023)Glennie, Adam, Leos-Barajas, Michelot, Photopoulou, and McClintock]{glennie2023}
Richard Glennie, Timo Adam, Vianey Leos-Barajas, Theo Michelot, Theoni Photopoulou, and Brett~T. McClintock.
\newblock Hidden markov models: pitfalls and opportunities in ecology.
\newblock \emph{Methods in Ecology and Evolution}, 14\penalty0 (1):\penalty0 43--56, 2023.

\bibitem[Holsclaw et~al.(2016)Holsclaw, Greene, Robertson, and Smyth]{holsclaw2016}
Tracy Holsclaw, Arthur~M. Greene, Andrew~W. Robertson, and Padhraic Smyth.
\newblock A bayesian hidden markov model of daily precipitation over south and east asia.
\newblock \emph{Hydrometeorology}, 17\penalty0 (1):\penalty0 3--25, 2016.

\bibitem[Patterson et~al.(2008)Patterson, Thomas, Wilcox, Ovaskainen, and Matthiopoulos]{patterson2008}
Toby~A Patterson, Len Thomas, Chris Wilcox, Otso Ovaskainen, and Jason Matthiopoulos.
\newblock State-space models of individual animal movement.
\newblock \emph{Trends in ecology \& evolution}, 23\penalty0 (2):\penalty0 87--94, 2008.

\bibitem[Thygesen et~al.(2009)Thygesen, Pedersen, and Madsen]{thygesen2009}
Uffe Thygesen, Martin Pedersen, and Henrik Madsen.
\newblock \emph{Geolocating Fish Using Hidden Markov Models and Data Storage Tags}, volume~9, pages 277--293.
\newblock 2009.

\bibitem[Turchin(1998)]{turchin1998}
Peter Turchin.
\newblock \emph{Quantitative Analysis of Movement: Measuring and Modeling Population Redistribution in Animals and Plants}.
\newblock Sinauer, 1998.

\bibitem[Prima et~al.(2018)Prima, Duchesne, Fortin, Rivest, and Fortin]{prima2018}
Marie-Caroline Prima, Thierry Duchesne, Andre Fortin, Louis-Paul Rivest, and Daniel Fortin.
\newblock Combining network theory and reaction-advection-diffusion modelling for predicting animal distribution in dynamic environments.
\newblock \emph{Methods in Ecology and Evolution}, 9\penalty0 (5):\penalty0 1221--1231, 2018.

\bibitem[McClintock et~al.(2012)McClintock, King, Thomas, Matthiopoulos, McConnel, and Morales]{mcclintock2012}
Brett~T. McClintock, Ruth King, Len Thomas, Jason Matthiopoulos, Bernie~J. McConnel, and Juan~M. Morales.
\newblock A general discrete-time modeling framework for animal movement using multistate random walks.
\newblock \emph{Ecological Monographs}, 82:\penalty0 335--349, 2012.

\bibitem[Fagan and Calabrese(2014)]{fagan2014}
William~F. Fagan and Justin~M. Calabrese.
\newblock The correlated rnadom walk and the rise of movement ecology.
\newblock \emph{The Bulletin of the Ecological Society of America}, 95:\penalty0 204--206, 2014.

\bibitem[Ahmed et~al.(2023)Ahmed, Bailey, and Bonsall]{ahmed2023}
Danish~A. Ahmed, Joseph~D. Bailey, and Michael~B. Bonsall.
\newblock On random walk models as a baseline for animal movement in three-dimensional space.
\newblock \emph{Ecological Modelling}, 475, 2023.

\bibitem[Turchin(1991)]{turchin1991}
Peter Turchin.
\newblock Translating foraging movements in heterogeneous environments into the spatial distribution of foragers.
\newblock \emph{Ecology}, 72\penalty0 (4):\penalty0 1253--1266, 1991.

\bibitem[Hedenstrom and Alerstam(1998)]{hedenstrom1998}
Anders Hedenstrom and Thomas Alerstam.
\newblock How fast can birds migrate?
\newblock \emph{Journal of Avian Biology}, 29\penalty0 (4):\penalty0 424--432, 1998.

\bibitem[McRae(2006)]{mcrae2006}
Brad~H McRae.
\newblock Isolation by resistance.
\newblock \emph{Evolution}, 60\penalty0 (8):\penalty0 1551--1561, 2006.

\bibitem[McRae and Beier(2007)]{mcrae2007}
Brad~H McRae and Paul Beier.
\newblock Circuit theory predicts gene flow in plant and animal populations.
\newblock \emph{Proceedings of the National Academy of Sciences}, 104\penalty0 (50):\penalty0 19885--19890, 2007.

\bibitem[McRae et~al.(2008)McRae, Dickson, Keitt, and Shah]{mcrae2008}
Brad~H. McRae, Brett~G. Dickson, Timothy~H. Keitt, and Viral~B. Shah.
\newblock Using circuit theory to model connectivity in ecology, evolution, and conservation.
\newblock \emph{Ecology}, 89\penalty0 (10):\penalty0 2712--2724, 2008.

\bibitem[Dickson et~al.(2019)Dickson, Albano, Anantharaman, Beier, Fargione, Graves, Gray, Hall, Lawler, Leonard, et~al.]{dickson2019}
Brett~G Dickson, Christine~M Albano, Ranjan Anantharaman, Paul Beier, Joe Fargione, Tabitha~A Graves, Miranda~E Gray, Kimberly~R Hall, Josh~J Lawler, Paul~B Leonard, et~al.
\newblock Circuit-theory applications to connectivity science and conservation.
\newblock \emph{Conservation biology}, 33\penalty0 (2):\penalty0 239--249, 2019.

\bibitem[Peterson et~al.(2019)Peterson, Hanks, Ver~Hoef, Hooten, and Fortin]{peterson2019}
Erin~E. Peterson, Ephraim~M. Hanks, Jay~M. Ver~Hoef, Mevin~B. Hooten, and Marie-Jos\'{e}e Fortin.
\newblock Spatially structured statistcal network models for landscape genetics.
\newblock \emph{Ecological Monographs}, 89\penalty0 (2):\penalty0 e01355, 2019.

\bibitem[Koen et~al.(2014)Koen, Bowman, Sadowski, and Walpole]{koen2014}
Erin~L Koen, Jeff Bowman, Carrie Sadowski, and Aaron~A Walpole.
\newblock Landscape connectivity for wildlife: development and validation of multispecies linkage maps.
\newblock \emph{Methods in Ecology and Evolution}, 5\penalty0 (7):\penalty0 626--633, 2014.

\bibitem[Grafius et~al.(2017)Grafius, Corstanje, Siriwardena, Plummer, and Harris]{grafius2017}
Darren~R Grafius, Ron Corstanje, Gavin~M Siriwardena, Kate~E Plummer, and Jim~A Harris.
\newblock A bird’s eye view: using circuit theory to study urban landscape connectivity for birds.
\newblock \emph{Landscape Ecology}, 32:\penalty0 1771--1787, 2017.

\bibitem[Chandra et~al.(1996)Chandra, Raghavan, Ruzzo, Smolensky, and Tiwari]{chandra1996}
Ashok~K. Chandra, Prabhakar Raghavan, Walter~L. Ruzzo, Roman Smolensky, and Prasoon Tiwari.
\newblock The electrical resistance of a graph captures its commute and cover times.
\newblock \emph{Computational complexity}, 6\penalty0 (4):\penalty0 312--340, 1996.

\bibitem[Hanks and Hooten(2013)]{hanks2013}
Ephraim~M. Hanks and Mevin~B. Hooten.
\newblock Circuit theory and model-based inference for landscape connectivity.
\newblock \emph{Journal of the American Statistical Association}, 108\penalty0 (501):\penalty0 22--33, 2013.

\bibitem[Thiele et~al.(2018)Thiele, Buchholz, and Schirmel]{thiele2018}
Jan Thiele, Sascha Buchholz, and Jens Schirmel.
\newblock Using resistance distance from circuit theory to model dispersal through habitat corridors.
\newblock \emph{Journal of Plant Ecology}, 11\penalty0 (3):\penalty0 385--393, 2018.

\bibitem[Christensen and Hoff(2024)]{christensen2024a}
Michael~F. Christensen and Peter~D. Hoff.
\newblock A flexible and interpretable spatial covariance model for data on graphs.
\newblock \emph{arXiv}, 2024.

\bibitem[Hankin(2006)]{hankin2006}
Robin~K.S. Hankin.
\newblock Resistor networks in r: Introducting the resistorarray package.
\newblock \emph{R News}, 6\penalty0 (2):\penalty0 52--54, 2006.

\bibitem[Klein and Randi\'{c}(1993)]{klein1993}
Douglas~J. Klein and Milan Randi\'{c}.
\newblock Resistance distance.
\newblock \emph{Journal of mathematical chemistry}, 12\penalty0 (1):\penalty0 81--95, 1993.

\bibitem[Paul(2001)]{paul2001}
Clayton~R Paul.
\newblock \emph{Fundamentals of electric circuit analysis}.
\newblock John Wiley \& Sons, 2001.

\bibitem[Hunt and Flaspohler(2020)]{warbler2020}
Peter~D. Hunt and D.~J. Flaspohler.
\newblock Yellow-rumped warbler (steophaga coronata).
\newblock In P.~G. Rodewald, editor, \emph{Birds of the World}. Cornell Lab of Ornithology, 2020.

\bibitem[Rising and Flood(2020)]{oriole2020}
J.~D. Rising and N.~J. Flood.
\newblock Baltimore oriole (icterus galbula).
\newblock In P.~G. Rodewald, editor, \emph{Birds of the World}. Cornell Lab of Ornithology, 2020.

\bibitem[Tryjanowski and Yosef(2002)]{tryjanowski2002}
Piotr Tryjanowski and Reuven Yosef.
\newblock Differences between the spring and autumn migration of the red-backed shrike lanius collurio: record from the eilat stopover (israel).
\newblock \emph{Acta Ornithologica}, 37\penalty0 (2):\penalty0 85--90, 2002.

\bibitem[Nilsson et~al.(2013)Nilsson, Klaassen, and Alerstam]{nilsson2013}
Cecilia Nilsson, Raymond~HG Klaassen, and Thomas Alerstam.
\newblock Differences in speed and duration of bird migration between spring and autumn.
\newblock \emph{The American Naturalist}, 181\penalty0 (6):\penalty0 837--845, 2013.

\bibitem[Sharma et~al.(2018)Sharma, Singh, Malik, Gupta, Rani, and Kumar]{sharma2018}
Aakansha Sharma, Devraj Singh, Shalie Malik, Neelu~Jain Gupta, Sangeeta Rani, and Vinod Kumar.
\newblock Difference in control between spring and autumn migration in birds: insight from seasonal changes in hypothalamic gene expression in captive buntings.
\newblock \emph{Proceedings of the Royal Society B}, 285\penalty0 (1885):\penalty0 20181531, 2018.

\bibitem[Kokko(1999)]{kokko1999}
Hanna Kokko.
\newblock Competition for early arrival in migratory birds.
\newblock \emph{Journal of Animal Ecology}, 68\penalty0 (5):\penalty0 940--950, 1999.

\bibitem[Christensen and Eidsvik(2024)]{christensen2024b}
Michael~F. Christensen and Jo~Eidsvik.
\newblock 2024.
\newblock \doi{arXiv}.

\bibitem[Marra et~al.(2005)Marra, Francis, Mulvihill, and Moore]{marra2005}
Peter~P Marra, Charles~M Francis, Robert~S Mulvihill, and Frank~R Moore.
\newblock The influence of climate on the timing and rate of spring bird migration.
\newblock \emph{Oecologia}, 142:\penalty0 307--315, 2005.

\bibitem[Covino et~al.(2020)Covino, Horton, and Morris]{covino2020}
Kristen~M Covino, Kyle~G Horton, and Sara~R Morris.
\newblock Seasonally specific changes in migration phenology across 50 years in the black-throated blue warbler.
\newblock \emph{The Auk}, 137\penalty0 (2), 2020.

\bibitem[Tottrup et~al.(2010)Tottrup, Rainio, Coppack, Lehikoinen, Rahbek, and Thorup]{tottrup2010}
Anders~P Tottrup, Kalle Rainio, Timothy Coppack, Esa Lehikoinen, Carsten Rahbek, and Kasper Thorup.
\newblock Local temperature fine-tunes the timing of spring migration in birds.
\newblock \emph{Integrative and Comparative Biology}, 50\penalty0 (3):\penalty0 293--304, 2010.

\end{thebibliography}

\end{document}